\documentclass[lettersize,journal]{IEEEtran}
\usepackage{amsmath,amsfonts}
\usepackage{algorithmic}
\usepackage{algorithm}
\usepackage{array}
\usepackage[caption=false,font=normalsize,labelfont=sf,textfont=sf]{subfig}
\usepackage{textcomp}
\usepackage{stfloats}
\usepackage{url}
\usepackage{verbatim}
\usepackage{graphicx}
\usepackage{cite}
\usepackage{booktabs}
\usepackage{amssymb}
\usepackage{pifont}%
\usepackage{hyperref}
\usepackage{bm}
\usepackage{threeparttable}
\usepackage{xcolor}
\let\origsubsubsection\subsubsection
\renewcommand{\subsubsection}{\par\vspace{0.5\baselineskip}\origsubsubsection}
\begin{document}

\title{Hypnos: A Dual-Phase End-to-End Platform\\ for Scalable Continual Learning at the Edge}

\author{Guilherme~Guedes,~\IEEEmembership{Graduate Student Member,~IEEE,}
        Marco~P.~E.~Apolinario,~\IEEEmembership{Member,~IEEE,}
        and~Charlotte~Frenkel,~\IEEEmembership{Senior Member,~IEEE}%
\thanks{This work has been submitted to the IEEE for possible publication. Copyright may be transferred without notice, after which this version may no longer be accessible.}%
\thanks{
This work was supported in part by the AdaptEdge project (Dutch Research Council, NWO Talent Programme — Veni, Grant 20267) and by the SynergAI project (Dutch Research Council and Microelectronics Department of TU Delft, AINed Fellowship research programme, file number NGF.1607.22.010).
G. Guedes, M. P. E. Apolinario, and C. Frenkel are with the Department of Microelectronics, Delft University of Technology, 2628 CD Delft, The Netherlands (e-mails: g.guedes@tudelft.nl; m.apolinario@tudelft.nl; c.frenkel@tudelft.nl). \textit{Corresponding author: G. Guedes.}}%
}

\maketitle
\begin{abstract}
Continual learning (CL) aims to enable the adaptation of neural networks to non-stationary data streams over time, by mitigating catastrophic forgetting.
In edge scenarios, CL algorithms offer a promising path to delivering robust, private, and adaptable devices.
Yet, despite promising results on large-scale datasets such as ImageNet, porting CL to the edge in the real world remains an open challenge for two main reasons: (i)~the online nature of operation, where sensor data arrives one sample at a time and the device must adapt to it on-the-fly, and (ii)~the low footprint demanded by edge devices, which makes running the traditional error backpropagation training algorithm prohibitive.
Inspired by the brain's wake and sleep cycles, the former for rapid adaptation with minimal resources and the latter for long-term consolidation, we introduce the open-source dual-phase platform \textit{Hypnos} for efficient CL at the edge. For the wake phase (\textit{e.g.}, daily user operation), Hypnos relies on a low-footprint system-on-chip with a RISC-V microcontroller unit (MCU) targeting efficient inference, with support for a lightweight gradient-free prototype update. For the sleep phase (\textit{e.g.}, device charge at night), where the device can have access to additional wall-plug energy and higher memory budgets, Hypnos features a smart charger that handles rehearsal-based gradient updates for consolidation.
Deployed on a ZCU104 field-programmable gate array (FPGA) and synthesized on a 28-nm process node, we benchmark on the CORe50 NICv2-391 protocol. Hypnos sets a new state-of-the-art accuracy of 78.6\%--81.6\% for edge CL systems with an on-device latency of 7.79\,s--10.54\,s per sample, depending on the activation precision. %
Additionally, it requires $3.8\times$ fewer learning steps and $22\times$ less learning compute than quantized latent replay at higher accuracy and comparable latency.
\end{abstract}

\begin{IEEEkeywords}
Continual learning (CL), convolutional neural networks (CNNs), image classification, non-stationary data streams, on-device learning, RISC-V, systems-on-a-chip (SoCs). 
\end{IEEEkeywords}

\section{Introduction}\label{sec:intro}

\begin{figure}[]
    \centering
    \includegraphics[width=\linewidth]{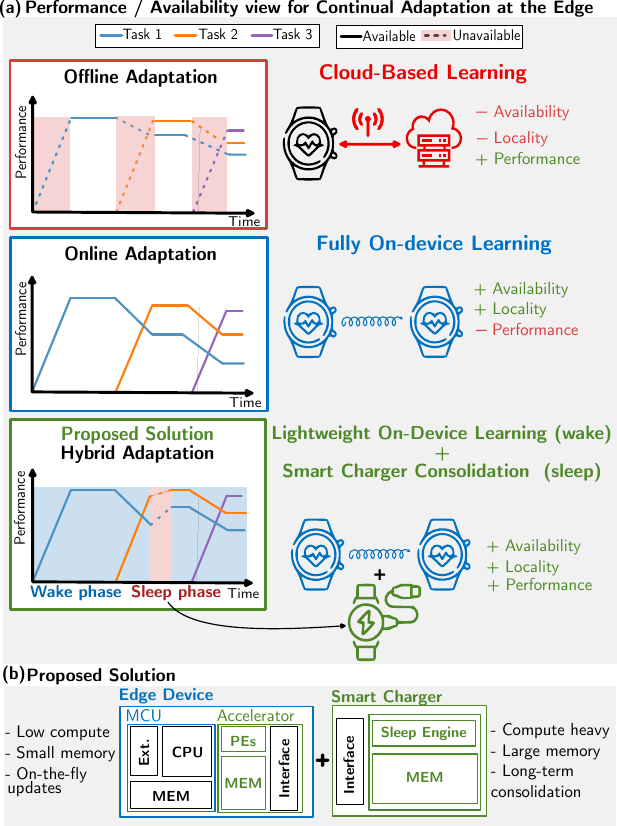}
    \caption{(a) Performance/availability view of continual adaptation at the edge. Top: offline adaptation, where consolidation is offloaded to the cloud and the device is unavailable (shaded) whenever new data must be learned. Middle: fully on-device learning keeps the device available with online adaptation, at the cost of increased catastrophic forgetting. Bottom: proposed hybrid scheme, lightweight on-device learning during daily use (wake) with consolidation on the smart charger (sleep), combining the availability of online methods with offline-like accuracy. (b) Proposed device--charger system realizing hybrid adaptation, combining (i)~a low-footprint edge device for online adaptation with (ii)~a smart charger augmented with additional compute and memory budgets for offline consolidation.}
    \label{fig:main}
\end{figure}

Battery-operated devices are becoming widespread, from wearable sensors that detect the onset of a health condition to smartwatches that learn their user's habits~\cite{murthyLearnLearnChip2022, verwimpContinualLearningApplications}.
These devices typically follow a \textit{train-then-deploy} scenario: a model is trained once, offline, on centrally collected data, and then deployed to the user device (Fig.~\ref{fig:main}(a), top).
Real-world deployments, however, can deviate from training conditions, as data distributions drift over time and devices can face environments and user
requirements unknown at training time~\cite{amodeiConcreteProblemsAI2016}.
Restoring accuracy by retraining in the cloud scales poorly to real-time needs: the energy and latency footprint of wireless communication for sending data and receiving an update~\cite{verwimpContinualLearningApplications} offsets the footprints of executing instructions locally~\cite{pottieWirelessIntegratedNetwork2000, shengEnergyEfficientCooperative2018}.

In contrast, on-device learning aims to remove this dependency by updating the model locally~\cite{ravagliaTinyMLPlatformOnDevice2021, blanken_chameleon_2025}.
However, efficient on-device learning is challenging as power and area budgets typical of edge devices, on the order of a few milliwatts and millimeters squared, severely cap on-device memory budgets~\cite{horowitz11ComputingsEnergy2014, ravagliaTinyMLPlatformOnDevice2021, blanken_chameleon_2025}.
Furthermore, representative edge scenarios typically rely on sensor data arriving online as a non-stationary stream, violating the independent and identically distributed~(iid) assumption. 
Yet, online learning from a stream of non-iid data is known to erase previously acquired knowledge (Fig.~\ref{fig:main}(a), middle), leading to \textit{catastrophic forgetting}~\cite{mccloskeyCatastrophicInterferenceConnectionist1989, frenchCatastrophicForgettingConnectionist2006}. 
Adapting to such streams is the goal of \textit{continual learning} (CL) algorithms, which must mitigate catastrophic forgetting to do so.
Among them, \textit{rehearsal-based methods} have been shown to scale best by replaying past data to emulate the iid assumption~\cite{vandevenThreeTypesIncremental2022, hayesOnlineContinualLearning2022}.
However, replay-based CL requires large memory budgets to store past data, as well as support for error backpropagation, thereby directly conflicting with edge requirements.\par

To resolve this conflict and make the best of offline and online methods, we build on the following observation: tight energy and area constraints at the edge only apply during the daily battery-operated phase of the device.
Indeed, edge devices typically have to be put to charge on a regular basis, a process during which limits induced by battery operation can be lifted by accessing a wall plug instead.
Therefore, we propose to tackle this challenge at the system level, with a low-footprint edge device embedding cheap online adaptation complemented by a \textit{smart charger} for memory- and compute-intensive offline consolidation (Figs.~\ref{fig:main}(a), bottom, and \ref{fig:main}(b)). 
This hybrid adaptation approach is akin to the brain's alternating wake and sleep phases, a promising strategy in CL~\cite{harunSIESTAEfficientOnline2023, vandevenBraininspiredReplayContinual2020a, verwimpContinualLearningApplications}.

In this work, we introduce \textit{Hypnos}, a dual-phase system for efficient yet scalable on-device CL, co-designed based on the SIESTA algorithm~\cite{harunSIESTAEfficientOnline2023} that introduces the use of cheap gradient-free adaptation (wake phase), which we link to daily user operation, combined with replay-based consolidation (sleep phase), which we link to the device being put to charge. Hypnos consists of
\begin{itemize}
    \item an open-source system-on-chip (SoC) targeting efficient inference at the edge, built on an X-HEEP RISC-V microcontroller unit (MCU)~\cite{machettiXHEEPOpenSourceConfigurable2025} with the Quadrilatero matrix co-processor~\cite{cammarataQuadrilateroRISCVProgrammable2025}, augmented with a compact precision-scalable wake-phase accelerator for the depth-wise convolutions, requantization, and prototype operations of online learning, accounting for 10.7\% of the device area,
    \item a smart charger with a 5.5-MB memory budget for replay-based consolidation and software-based support for error backpropagation.
\end{itemize}
On the CORe50~\cite{lomonacoCORe50NewDataset} NICv2-391 protocol, a full field-programmable gate array (FPGA) demonstration of Hypnos reaches an accuracy of 78.6\,\% at an on-device latency of 7.79\,s per sample, 3.4\,pp above the best prior edge CL system at comparable latency~\cite{ravagliaTinyMLPlatformOnDevice2021}, while learning fully online and on-device. A variant with INT16 activations trades latency for accuracy, reaching 81.6\,\% at 10.54\,s per sample. Hypnos requires 3.8× fewer learning steps and 22× less learning compute than quantized latent replay at higher accuracy. 
Synthesis in a 28-nm node shows that the complete edge device occupies 9.43\,mm$^2$ and consumes 33.3\,mW at 375\,MHz, with the wake-phase accelerator accounting for only 10.7\,\% of the area.  The \textit{Hypnos} system is available in open source at \url{https://github.com/cogsys-tudelft/hypnos}.
 
The remainder of this paper is structured as follows.
Section~\ref{sec:bg} provides background on CL and its edge deployments, as well as low-footprint techniques.
Section~\ref{sec:algo} introduces our framework for dual-phase CL systems.
Section~\ref{sec:codesign} describes the implementation of the framework through system co-design across the algorithm, the wake-phase accelerator, and the smart charger.
Finally, Section~\ref{sec:results} evaluates Hypnos based on FPGA and synthesis results, an accelerator ablation, and a comparison with the state of the art on CORe50, while Section~\ref{sec:conclusion} concludes the paper.

\section{Background}\label{sec:bg}
In this section, we formally define CL and outline its taxonomy. 
We then elaborate on the low-footprint ingredients underpinning dual-phase algorithms: prototype-based learning, latent replay, quantization, and compression. Finally, we review the hardware platforms that demonstrate them at the edge, closing on the gap this work addresses.
\subsection{Continual Learning (CL)}

Deep learning assumes that data comes from a stationary distribution and is fully available at training time~\cite{lecunDeepLearning2015, parisiLifelongLearningSpatiotemporal2018}, but real-world data streams violate both assumptions.
Formally, the environment is modeled as an ordered sequence of tasks $\mathcal{T} = (T_1, T_2, \dots, T_N)$, where each task $T_i$ presents input--label pairs $(x, y)$ drawn from their own distribution. 
In the class-incremental learning (CIL) scenarios considered in this work, label sets are additionally disjoint across tasks, \textit{i.e.}, $\mathcal{Y}_i \cap \mathcal{Y}_j = \emptyset$ for $i \neq j$~\cite{vandevenThreeTypesIncremental2022}. 
Continual learning is the problem of optimizing a single set of model parameters $\theta$ over this sequence such that performance on earlier tasks is preserved.
Naively training $\theta$ on $T_{i+1}$ erases what was learned on $T_i$, a behavior called \textit{catastrophic forgetting} shown at various degrees in Fig.~\ref{fig:main}(a).
Among the standard CL scenarios~\cite{vandevenThreeTypesIncremental2022}, this work targets \textit{online CIL}~\cite{hayesOnlineContinualLearning2022}, representative of edge scenarios, which adds two constraints: (i) samples arrive in a single-pass stream and are processed once, with no multi-epoch training; and (ii) task identities are hidden at inference, so the model must discriminate among all classes seen so far.

Existing CL algorithms fall into three main categories~\cite{vandevenThreeTypesIncremental2022}, which differ not only in how they mitigate catastrophic forgetting but also in how amenable they are to low-footprint execution.
First, \textit{architectural} approaches expand the network's capacity to accommodate new knowledge~\cite{rusuProgressiveNeuralNetworks2022, vonoswaldContinualLearningHypernetworks2022, aljundiExpertGateLifelong2017, yuBoostingContinualLearning2024}. 
Second, \textit{regularization} approaches constrain parameter updates based on parameter importance~\cite{kirkpatrickOvercomingCatastrophicForgetting2017, zenkeContinualLearningSynaptic2017, liLearningForgetting2018, aljundiMemoryAwareSynapses2018, chaudhryRiemannianWalkIncremental2018, magistriElasticFeatureConsolidation2024} or orthogonal gradient projections~\cite{lopez-pazGradientEpisodicMemory2017, huGradientEpisodicMemory2020}. 
Lastly, \textit{rehearsal} approaches retain a subset of past samples in a memory buffer and revisit them during training~\cite{chaudhryTinyEpisodicMemories2019, buzzegaDarkExperienceGeneral2020, shinContinualLearningDeep2017, wangOnlineContrastiveContinual2025}.
Among these, rehearsal has proven the most robust and scalable~\cite{vandevenThreeTypesIncremental2022}, owing to its ability to emulate an iid data stream, at the expense of additional compute for every replayed sample and a memory buffer that grows with the number of retained classes. 
These memory accesses, which are commonly off-chip, cost orders of magnitude more energy than the compute~\cite{horowitz11ComputingsEnergy2014}, increasing not only latency but also power consumption.

\subsection{Key Techniques for Low-Footprint CL}\label{ssec:low_footprint_cl}

In this subsection, we highlight several key techniques that have been used toward low-footprint CL and edge deployment, the ingredients from which dual-phase algorithms are composed.

\subsubsection{Prototype-based learning}
\begin{figure}
    \centering
    \includegraphics[width=\linewidth]{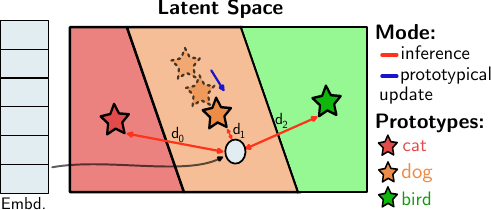}
    \caption{Prototype-based network in the latent space. Stars represent the class prototypes $\bm{\mu}_k$; the circle is the embedding $\mathbf{z}_t$ of an incoming sample. During inference (red), each class is scored by the distance $d_k$ between $\mathbf{z}_t$ and its prototype. On a labeled sample, the matching prototype moves as a running mean via the update of Eq.~\eqref{eq:ncm}, represented by the blue line and the moving prototype.}
    \label{fig:ncm}
\end{figure}

Within architectural approaches to CL, one of the lightest forms of expansion is \textit{prototype-based learning}~\cite{snellPrototypicalNetworksFewshot2017} (Fig.~\ref{fig:ncm}): each class $k$ is represented by a single vector in the latent space of a frozen feature extractor, its \textit{prototype}~$\bm{\mu}_k$, obtained as the running mean of the features of all samples of that class seen so far. 
Upon receiving the latent representation~$\mathbf{z}_t$ of a sample belonging to class~$k$, the prototype is updated as
  \begin{equation}
    \bm{\mu}_k \leftarrow \frac{c_k\,\bm{\mu}_k + \mathbf{z}_t}{c_k + 1},
    \label{eq:ncm}
  \end{equation}
  where $c_k$ counts the samples of class~$k$ seen so far. 
During inference, the classifier scores each of the $N$~classes seen so far by the similarity between a new sample's latent representation~$\mathbf{z}$ and each prototype:
\begin{equation}
  p_k = \frac{\exp\left(\mathrm{sim}(\mathbf{z},\,\bm{\mu}_k)\right)}
             {\sum_{n=1}^{N} \exp\left(\mathrm{sim}(\mathbf{z},\,\bm{\mu}_n)\right)},
  \label{eq:proto-clf}
\end{equation}
where $\mathrm{sim}(\cdot,\cdot)$ is a similarity metric (\textit{e.g.}, the cosine similarity~\cite{harunSIESTAEfficientOnline2023}), and the class with the highest probability~$p_k$ is selected as the prediction. 
Neither operation requires gradient computation, making the approach light enough for online, per-sample adaptation. 
Moreover, using the squared L2 distance as similarity metric, the input norm~$\lVert\mathbf{z}\rVert^2$ is a constant offset common to all classes, so a nearest-prototype scoring reduces to the affine form
\begin{equation}
  \mathbf{y} = \mathbf{z}\mathbf{W} + \mathbf{b},
  \qquad
  \mathbf{W} = -2\,[\bm{\mu}_1, \dots, \bm{\mu}_N],
  \qquad
  b_i = \lVert\bm{\mu}_i\rVert^2,
  \label{eq:fc-trick}
\end{equation}
where $\mathbf{y}$ collects the $N$ class scores, the weight matrix stacks the prototypes, and the biases carry their norms. Adding or updating a class amounts to appending or updating one row of a fully-connected (FC) layer~\cite{blanken_chameleon_2025}.

\subsubsection{Latent replay}
\begin{figure}
    \centering
    \includegraphics[width=\linewidth]{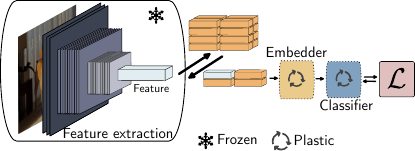}
    \caption{Latent replay. The network is split at the \textit{replay layer}: the frozen extractor below the split runs once per sample and its activations are stored in the latent sample buffer. Only the layers above the split remain plastic and are trained by gradient descent, interleaving buffered activations with new ones.} 
    \label{fig:latent-replay}
\end{figure}

Instead of storing raw samples as in traditional rehearsal~\cite{chaudhryTinyEpisodicMemories2019}, \textit{latent
replay}~\cite{pellegriniLatentReplayRealTime2020} buffers the latent activations of an intermediate layer, and only the layers above it remain plastic (Fig.~\ref{fig:latent-replay}). 
This reduces both the compute and the memory cost of consolidation, and introduces a trade-off in the choice of the \textit{replay layer}, as its depth balances compression against representational richness~\cite{ravagliaTinyMLPlatformOnDevice2021}.

\subsubsection{Quantization}
In the uniform affine scheme~\cite{jacobQuantizationTrainingNeural2018, nagelWhitePaperNeural2021}, integer quantization represents a real-valued element $x$ by an integer $x_q$, a scale $s_x$, and a zero-point $z_x$ such that $x \approx s_x\,(x_q - z_x)$, with $x_q$ obtained as
\begin{equation}\label{eq:quant}
    x_q = \mathrm{clamp}\left(\left\lfloor \frac{x}{s_x}
    \right\rceil + z_x,\, -2^{b-1},\, 2^{b-1}-1\right),
\end{equation}
where $b$ is the bitwidth and $\lfloor\cdot\rceil$ denotes rounding to the nearest integer. Weights are commonly quantized symmetrically ($z_w = 0$).
A layer then executes entirely in integers: the products $x_q\,w_q$ accumulate at full precision into an accumulator $a$, and a \textit{requantization} step returns the accumulator to the $b$-bit activation grid as
\begin{equation}\label{eq:requant}
    y_q = \mathrm{clamp}\left(\left\lfloor a \cdot r \right\rceil
    + z_y,\, -2^{b-1},\, 2^{b-1}-1\right),
\end{equation}
where $r = s_x s_w / s_y \approx M \cdot 2^{-n}$ combines the scales of the two operands and the output. This is approximated by an integer mantissa $M$ and a right-shift $n$, and $z_y$ is the output zero-point, removing the need for floating-point (FP) multiplication or division.
Mixed-precision variants assign different bitwidths per tensor, trading accuracy against footprint~\cite{nagelWhitePaperNeural2021}.

\subsubsection{Compression} \label{sssec:compression}
\begin{figure}
    \centering
    \includegraphics[width=\linewidth]{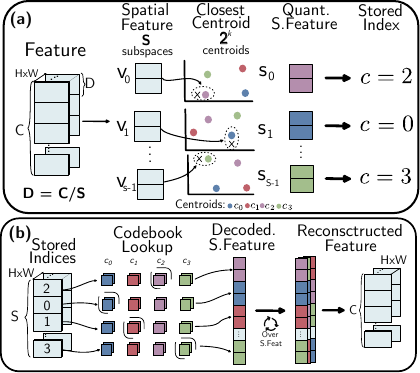}
    \caption{Product quantization. \textbf{(a)} Encoding: the feature vector is split channel-wise into $S$ sub-vectors of dimension $D = C/S$. Each sub-vector is then mapped to the nearest of the $2^k$ centroids of its subspace codebook (Algorithm~\ref{alg:pq}) and only its $k$-bit index is stored. \textbf{(b)} Decoding: each stored index selects its centroid from the codebook, a table lookup which reconstructs the sub-vectors. Repeating it over all spatial positions reconstructs the full feature map.}
    \label{fig:pq}
\end{figure}

Even in latent form, the replay buffer dominates the memory footprint of rehearsal.
Linear quantization of the stored activations reduces it by only a few times, a gain that varies widely with the benchmark and the replay layer: on CORe50, the quantized latent replay buffer of~\cite{ravagliaTinyMLPlatformOnDevice2021} still occupies 3 to 98\,MB depending on the replay layer, one to two orders of magnitude beyond the on-chip memory of an MCU.
This calls for compression techniques.
\textit{Product quantization} (PQ)~\cite{jegouProductQuantizationNearest2011} is a technique from approximate nearest-neighbor search and database storage~\cite{johnsonBillionscaleSimilaritySearch2017}.
A feature vector $\mathbf{z} \in \mathbb{R}^{C}$ is divided into $S$ disjoint sub-vectors $[\mathbf{z}_1, \dots, \mathbf{z}_S]$ of dimension $D = C/S$ (Fig.~\ref{fig:pq}(a)), and each subspace $m$ is quantized independently by a codebook of $2^{k}$ centroids $\{\mathbf{s}_m^{(j)}\}_{j=1}^{2^k}$, indexed by $j$ and learned offline with the k-means algorithm.
A sample is encoded by storing, for each subspace, only the index of its nearest centroid, as summarized in Algorithm~\ref{alg:pq}.
Decoding corresponds to a table lookup that reconstructs each sub-vector from its centroid (Fig.~\ref{fig:pq}(b)).
\begin{algorithm}[t]
\caption{PQ encoding of a feature vector}
\label{alg:pq}
\begin{algorithmic}[1]
\REQUIRE $\mathbf{z} \in \mathbb{R}^{C}$; codebooks
$\{\mathbf{s}_m^{(j)}\}_{j=1}^{2^k}$ for $m = 1, \dots, S$
\ENSURE code $(i_1, \dots, i_S)$, $k$ bits per index
\STATE split $\mathbf{z}$ into $S$ sub-vectors $[\mathbf{z}_1, \dots, \mathbf{z}_S]$,
each of dimension $D = C/S$
\FOR{$m = 1$ \TO $S$}
    \STATE $i_m \gets \arg\min_{j}\,
    \lVert \mathbf{z}_m - \mathbf{s}_m^{(j)} \rVert^{2}$
    \COMMENT{nearest centroid in subspace $m$}
\ENDFOR
\RETURN $(i_1, \dots, i_S)$
\end{algorithmic}
\end{algorithm}

PQ thereby decouples the storage cost from the channel dimension.
While the linear quantization compression ratio is bounded by how far the bitwidth can shrink before quantization noise dominates, PQ adds the number of subspaces $S$ as a second degree of freedom, an advantage for latent replay, whose features come from the middle layers where $C$ is large.
PQ gives a compression ratio of $C\,b/(S\,k)$ for a $C$-channel $b$-bit vector encoded as $S$ $k$-bit indices, an order of magnitude beyond linear quantization (see Section~\ref{subsec:quant}).
In terms of overhead, the codebooks occupy $2^{k} \cdot C$ values of storage, and every incoming sample adds $S$ nearest-centroid searches, a compute overhead relative to the element-wise operations of linear quantization.

\subsection{CL Deployment on Edge Hardware}
Each of the preceding ingredients has been demonstrated at the edge, in hardware, or as algorithms designed under edge constraints. This subsection reviews these demonstrations, which together highlight the gap this work addresses. \par

The prototype-based classifier, in the affine form of Eq.~\eqref{eq:fc-trick}, was demonstrated in hardware by the Chameleon accelerator~\cite{blanken_chameleon_2025}, which learns the 1623 handwritten character classes of Omniglot~\cite{lakeHumanlevelConceptLearning2015} online and fully on-chip. 
Latent replay was demonstrated on-device by the quantized latent replay platform (QLR-CL) of~\cite{ravagliaTinyMLPlatformOnDevice2021}, built on the VegaSoC~\cite{rossiVegaTenCoreSoC2022} MCU, the first to solve CORe50 on-device by quantizing the latent replay buffer.
Aggarwal \textit{et al.}~\cite{aggarwalChameleonDualMemory2024} augment it with a dual-memory system operating at two learning time scales, a fast adaptive memory and a slow consolidating one, deployed on both an FPGA and an embedded graphics processing unit (GPU).
Both memories, however, are consolidated by gradient-based replay on the device itself, so the device must host error backpropagation and the full replay buffer at all times, with computational and memory resources beyond edge MCU budgets.
Quantization underpins both of these demonstrations, from the quantized backbone and latent replay buffer of QLR-CL~\cite{ravagliaTinyMLPlatformOnDevice2021} to the multiplier-free integer arithmetic of Chameleon~\cite{blanken_chameleon_2025}, although both confine it to the frozen portion of the network.
The 22-nm accelerator of~\cite{suzuki22nmContinualLearning2026} extends quantization to the learning path itself, co-designing a sparse update rule with a mixed-precision INT4/INT8 tensor engine that updates only 1\,\% of the
weights per task, although its consolidation remains an offline, batched process. 
CL has also been explored in neuromorphic settings~\cite{huoANPG28nm104pJ2023, pesActiveDendritesEnable}, although only on small-scale low-dimensional tasks.

Every platform that reaches real-image workloads~\cite{ravagliaTinyMLPlatformOnDevice2021, aggarwalChameleonDualMemory2024, suzuki22nmContinualLearning2026} consolidates offline, in batched processes, whereas fully on-chip online learning has been shown only at Omniglot scale~\cite{blanken_chameleon_2025}.
A platform that learns online, fully on-chip, from real-image streams is thus an open challenge.

\section{Dual-Phase CL Framework}\label{sec:algo}
\begin{figure*}[t]
    \centering
    \includegraphics[width=0.75\textwidth]{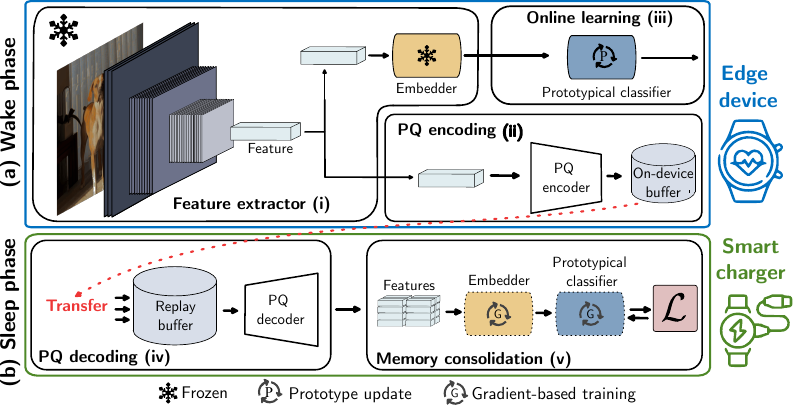}
    \vspace*{-3mm}\caption{System dual-phase learning framework, consisting of two phases. \textbf{(a)} Wake phase: a frozen feature extractor generates features that are compressed to be used during sleep (ii), and used for online learning (iii) using the running-mean update of Eq.~\eqref{eq:ncm}. The wake phase is handled by an MCU minimally extended with a daily sample buffer and prototypical learning. \textbf{(b)} Sleep phase: the online buffer is moved onto a larger buffer and the samples are decoded (iv) and fed to the same embedder and classifier, now trained using standard gradient-based methods (v). The sleep phase is handled via a smart charger, benefiting from extended memory and power budgets.}
    \label{fig:algorithm}
\end{figure*}
Novel dual-phase algorithms such as SIESTA~\cite{harunSIESTAEfficientOnline2023} are best understood as a combination of the techniques introduced in Section~\ref{ssec:low_footprint_cl}: the online \textit{wake phase} is carried out based on cheap architectural CL techniques (i.e.~prototype-based learning), while the offline \textit{sleep phase} relies on higher-footprint rehearsal approaches (i.e.~latent replay) for consolidation against catastrophic forgetting. The latter is accompanied by techniques to mitigate the memory overhead, such as PQ. Our proposal is to leverage this dual-phase algorithmic insight into a \textit{system framework} expanding it with a hardware view, as shown in Fig.~\ref{fig:algorithm}, consisting of two main parts.

First, the \textit{wake phase} is shown in Fig.~\ref{fig:algorithm}(a), as follows:
\begin{itemize}
    \item \textit{Algorithm} -- A frozen backbone, consisting of the feature extractor and the embedder, processes each incoming sample (i). Its features are used twice: the activations at the latent split are encoded with PQ into a small buffer for consolidation during the next sleep phase (ii), while the embedding drives online learning through the running-mean prototype update of Eq.~\eqref{eq:ncm} (iii).
    \item \textit{Hardware} -- The wake phase can be efficiently supported by an inference accelerator that primarily aims at efficiently executing the backbone feature extractor and embedder. It is minimally extended with an accelerator for prototypical learning and a small on-device buffer for storing, in compressed form, new samples encountered during the day.    
\end{itemize}\vspace*{1mm}
    
Second, the \textit{sleep phase} is shown in Fig.~\ref{fig:algorithm}(b), as follows:
\begin{itemize}
    \item \textit{Algorithm} -- The contents of small wake-phase buffer are transferred to the larger replay buffer maintained across wake/sleep cycles. Contents from this buffer are then sampled and decoded from PQ (iv). The samples are then used for memory consolidation (v) with latent replay, where gradient-based methods are used to retrain the full network head, including the embedder.
    \item \textit{Hardware} -- The sleep phase requires large memory and computational budgets, the former for the replay buffer and the latter for gradient-based learning. A smart charger benefiting from wall-plug power and endowed with extended memory allows fully offloading this overhead from the edge device.
\end{itemize}\vspace*{1mm}

The proposed \textit{device--charger system} allows keeping the edge device within adequate power and memory budgets (mW, MB), while alleviating the accuracy challenges resulting from catastrophic forgetting. Importantly, this system has a \textit{key tradeoff between plasticity and memory overhead} based on where a baseline neural network is split into a frozen backbone feature extractor and an embedder adapted during sleep. Indeed, a shallower backbone and a deeper embedder will typically increase adaptation capacity during sleep, at the expense of larger replay buffers due to higher sample dimensionality. This choice thus couples the accuracy, the buffer size, and the sleep-phase compute, and is further conditioned on the precision the deployed network is quantized to. This needs to be tackled through adequate hardware/algorithm co-design, which we are carrying out in the next section.

\section{System Co-Design}\label{sec:codesign}

Based on the framework introduced in Section~\ref{sec:algo}, we now set the use case and co-design the dual-phase algorithm and the hardware system that executes it, targeting a fully on-device wake phase. The workload and experimental setup are first introduced in Section~\ref{subsec:setup}, and we then proceed with hardware-aware algorithm exploration (Section~\ref{subsec:quant}) and device--charger hardware platform design (Section~\ref{subsec:platform}).

\subsection{Use Case and Experimental Setup}\label{subsec:setup}

Targeting a real-world scenario for CL, we adopt the CORe50 benchmark (NICv2-391 protocol), where 50 objects (classes) are recorded across several sessions, i.e.~short videos of 300 frames representing distinct environments that vary background, illumination, pose, and occlusion. Of the eleven available sessions for each class, three are held out for testing. 391 batches are then constructed as follows: an initial batch contains 10 class-sessions, which is followed by 390 incremental batches of one class-session each, where a class-session thus contains 300 images. These small, temporally correlated, single-class batches approximate an online, non-iid data stream better than a traditional split dataset scenario, making it a common test for edge CL~\cite{ravagliaTinyMLPlatformOnDevice2021,aggarwalChameleonDualMemory2024}. 

As the backbone, we adopt MobileNetV3-Large~\cite{howardSearchingMobileNetV32019}, shown to be a strong low-footprint backbone for image processing~\cite{harunSIESTAEfficientOnline2023}. Compared to the MobileNetV1 backbone of prior edge-CL platforms~\cite{ravagliaTinyMLPlatformOnDevice2021, aggarwalChameleonDualMemory2024}, where the operator set consists of dense point-wise (PW) and depth-wise (DW) convolutions followed by ReLU, MobileNetV3 adds two operations. The first one is the squeeze-and-excitation (SE) block~\cite{hu2018squeeze}, in which two small fully-connected (FC) layers compute a per-channel gate in $[0,1]$ from the spatially averaged feature map and rescale each channel accordingly. The second is the \textit{hard-swish} activation, $x \cdot \mathrm{hsig}(x)$ with $\mathrm{hsig}(x) = \mathrm{ReLU_6}(x + 3)/6$, which replaces ReLU in the deeper layers.

\subsection{Hardware-Aware Algorithmic Exploration}\label{subsec:quant}

We are now focusing on the key tradeoff between plasticity and memory overhead, which we outlined at the end of Section~\ref{sec:algo}. Given the large search space, we explore this tradeoff in two steps. We will first do an exploration of the split layer without quantization, from which we will narrow down the search to find suitable quantized configurations.

\subsubsection*{Latent replay split}\label{subsec:split}
\begin{figure}[t]
    \centering
    \includegraphics[width=\linewidth]{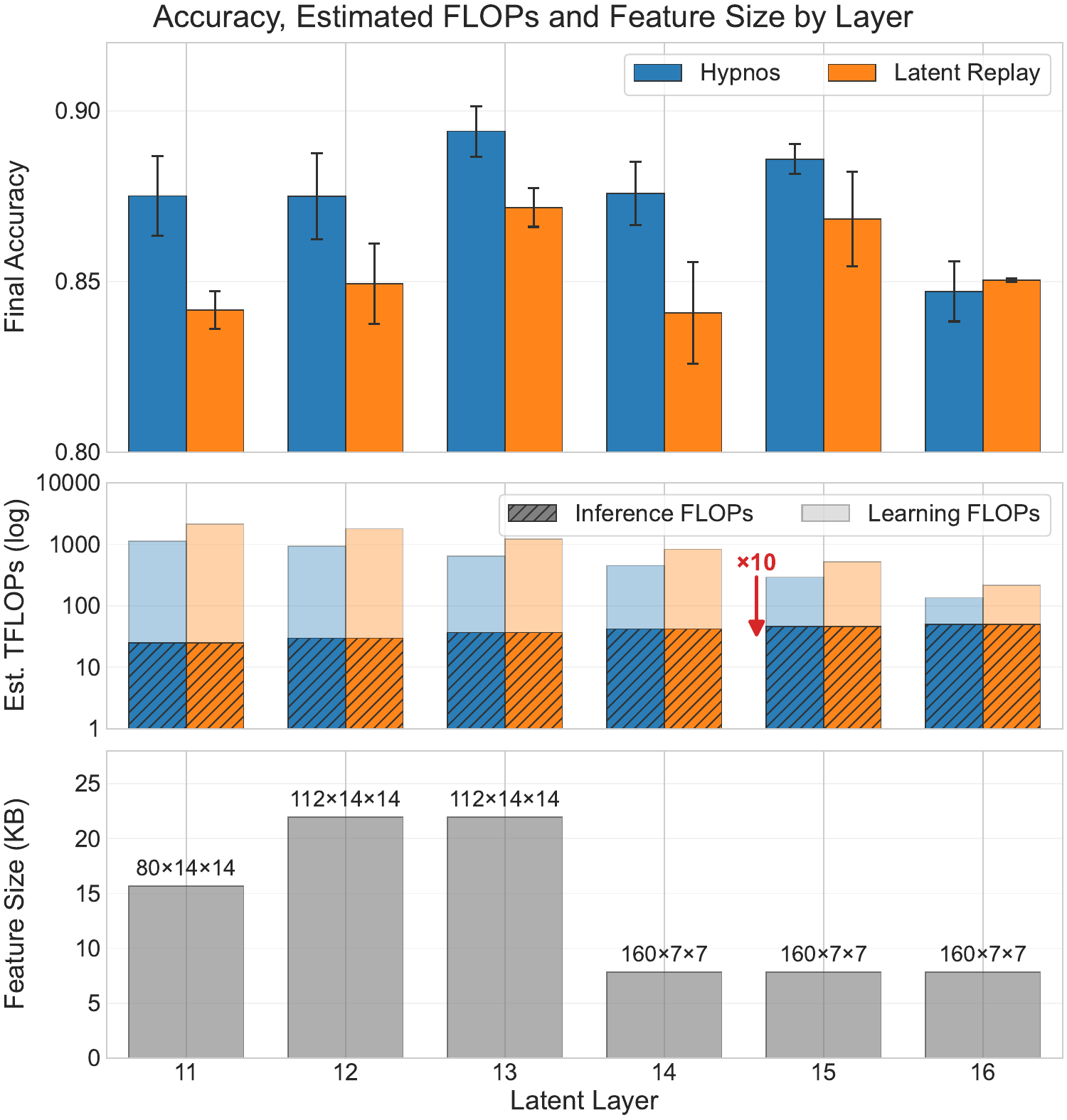}
    \caption{Latent-layer sweep of SIESTA and LR at FP32 precision on the NICv2-391 protocol of CORe50 with a replay buffer of 7000 samples. Top: final accuracy, mean and standard deviation over five runs. Middle: estimated compute over the full protocol, split into inference FLOPs, the forward passes through the frozen layers, and learning FLOPs, the gradient-based updates of the plastic layers, which SIESTA executes during the sleep phase and LR upon every incoming batch. Bottom: size of one latent sample ($C \times H \times W$) stored in the replay buffer, identical for both algorithms.}
    \label{fig:flop_comp}
\end{figure}

Taking our MobileNetV3 network, we will first assess at which layer it is suitable to split it between a frozen feature extraction backbone and a head consisting of an embedder and prototypical classifier (Fig.~\ref{fig:algorithm}), defining it as the \textit{latent layer}. We thus compare the accuracy--memory--compute tradeoff of the SIESTA and latent replay (LR) algorithms while sweeping the latent layer, as shown in Fig.~\ref{fig:flop_comp}, from which two main conclusions can be drawn. 
First, SIESTA matches or outperforms LR in accuracy at every split, at a lower compute budget. 
Second, the latent layer trades accuracy against sleep-phase compute and sample size (Fig.~\ref{fig:flop_comp}, middle and bottom). Layers 14--16 share the smallest sample footprint, and among them layer 15 gives the highest accuracy. We therefore adopt the latter as the latent layer, since the additional learning compute it entails relative to layer 16 is paid during the sleep phase only, where the charger's additional energy budget allows affording the compute overhead.

\subsubsection*{Quantized deployment}\label{subsec:precision}
\begin{table}[t]
\centering
\caption{Accuracy (\%) of candidate latent splits on CORe50
\mbox{NICv2-391} versus activation precision, with and without
PQ-compressed replay, as mean $\pm$ standard deviation over three runs.}
\label{tab:quant_ablation}
\footnotesize
\begin{tabular}{@{}lcccc@{}}
\toprule
 & \multicolumn{2}{c}{Layer 15} & \multicolumn{2}{c}{Layer 16} \\
\cmidrule(lr){2-3}\cmidrule(lr){4-5}
Activations & w/o PQ & w/ PQ & w/o PQ & w/ PQ \\ \midrule
FP32  & $83.2 \pm 1.1$ & $80.8 \pm 1.7$ & $82.6 \pm 1.3$ & $80.4 \pm 0.9$ \\
INT16 & $83.2 \pm 0.7$ & $\mathbf{81.6 \pm 2.1}$ & $82.5 \pm 1.7$ & $81.3 \pm 0.7$ \\
INT8  & $79.3 \pm 1.0$ & $\mathbf{78.6 \pm 1.2}$ & $79.9 \pm 1.1$ & $77.8 \pm 1.4$ \\ \bottomrule
\end{tabular}
\end{table}

Keeping the wake phase fully on-chip demands that the backbone, the online buffer, and the prototypes are kept in on-chip memory. We thus quantize the entire wake phase to mixed integer precision, following the uniform affine scheme of Eqs.~\eqref{eq:quant}--\eqref{eq:requant}, and adopt PQ for compressing sample storage requirements.

Weight quantization is subject to strict constraints in order to keep all network parameters on-device. For backbone convolutions, we adopt 8-bit weights with per-channel scales, as MobileNet-class networks have been shown to require them based on their point-wise kernel distributions~\cite{nagelWhitePaperNeural2021}. The SE FC layers, which account for 51\,\% of the feature-extractor parameters, are reduced further to 4-bit weights, as FC layers tolerate lower precision than convolutional kernels~\cite{hanDeepCompressionCompressing2016}. Batch-normalization layers are folded into their preceding convolution weights~\cite{krishnamoorthiQuantizingDeepConvolutional2018, jacobQuantizationTrainingNeural2018}. Under this scheme, the frozen backbone can be reduced to 3.7\,MB.

In order to reduce buffering requirements, the PQ scheme of Section~\ref{sssec:compression} can be adopted. Given a sample of spatial dimensions $W \times H$ with $C$ channels at the chosen latent layer, we treat each spatial location as an independent vector, reshaping the feature map into $WH$ vectors of dimension $C$. It is then compressed via PQ into $WH \cdot S \cdot k$ bits, against the $WH \cdot C \cdot b$ bits of its raw form, which for $S = 16$ subspaces of $k = 8$ bits leads to a $40\times$ compression for layer-15 samples (from 31.3\,kB to 784\,B). We size the replay buffer at 7000 samples (5.48\,MB) to stay within the 5.5-MB memory budget of the charger. The PQ centroid sets are fitted on features produced by the quantized backbone offline.

Based on the parameters above, Table~\ref{tab:quant_ablation} compares the candidate splits for FP32, INT16, and INT8 activations, each with and without PQ. Activations use one scale and zero-point per tensor, so the requantization factor $r$ is applied per channel. Prototypes are always kept at 16 bits as the per-sample correction of the running mean of Eq.~\eqref{eq:ncm} shrinks as a class accumulates samples, requiring more headroom. Three observations follow.
First, once the network is quantized, the split between layer 15 and layer 16 matters less accuracy-wise, as the two layers stay within one standard deviation of each other at every precision.
Second, the accuracy is stable from FP32 down to INT16, which differ by at most 0.9\,pp, comparable to the standard deviation over runs, and the drop concentrates at INT8, 3.0\,pp at layer 15 with PQ, revealing a memory--accuracy tradeoff as INT16 requires a larger scratch memory.
Third, PQ with INT8 centroids costs 0.7 to 2.4\,pp across precisions, so the $40\times$ buffer compression of Section~\ref{subsec:split} does not amplify the quantization loss.

\begin{figure*}
    \centering
    \includegraphics[width=\linewidth]{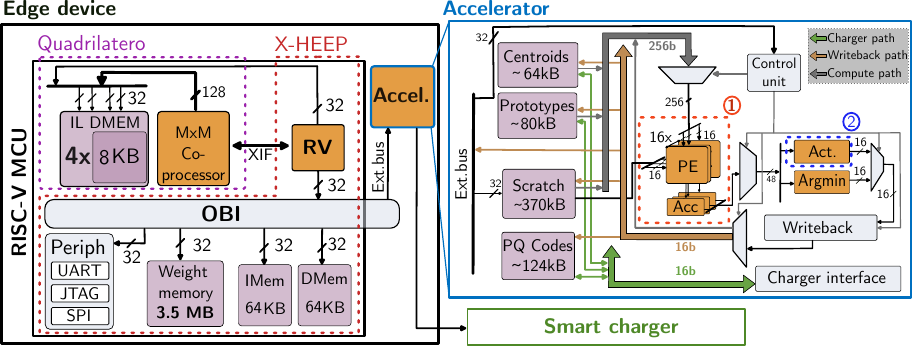}
    \caption{Full system block diagram. The system consists of an edge device and a smart charger. The edge device contains a RISC-V MCU implemented through the X-HEEP platform with a dedicated matrix co-processor. The accelerator consists of four memories, a 16-lane PE array, and the post-MAC operation blocks. The scratch memory holds the input and output activations of the layer being executed, the centroid and prototype memories hold the PQ codebooks and the class prototypes, and the PQ-code memory stores the encoded samples of the online buffer. The {\color{red}{\raisebox{.5pt}{\textcircled{\raisebox{-0.9pt}{1}}} PE array}} is a flexible PE-array that supports $16\times$INT16, $32\times$INT8 or $64\times$INT4 multiplications in parallel with 48-bit wide accumulators. The {\color{blue}{\raisebox{.5pt}{\textcircled{\raisebox{-0.9pt}{2}}} post-op blocks}}, Act, Argmax, and writeback, are used depending on the instruction sent to the accelerator and are responsible for applying an activation function, determining the closest prototype/centroid and whether to write back the results or not, respectively.}
    \label{fig:system}
\end{figure*}

\subsection{Dual-Phase Hardware Platform Design}\label{subsec:platform}

We introduce \textit{Hypnos}, a hardware platform for dual-phase continual learning, as shown in Fig.~\ref{fig:system}. The edge device is a system-on-chip (SoC) built around a \textit{RISC-V MCU}, embedding
\begin{itemize}
    \item a \textit{matrix co-processor} to accelerate the dense PW convolutions of the frozen feature extraction backbone,
    \item a compact \textit{wake-phase accelerator} to handle the low-density operations that neither the scalar core of the MCU nor the matrix co-processor executes efficiently: the DW convolutions, the per-channel requantization of the frozen backbone, the distance and update primitives of the classifier, and PQ encoding.
\end{itemize}

This edge device is complemented by a \textit{smart charger} that hosts the sleep phase. In this section, we outline these main blocks and monitor key design choices.

\subsubsection*{MCU Baseline}
In order to build a SoC based on a configurable, open-source RISC-V MCU, we adopt the X-HEEP platform~\cite{machettiXHEEPOpenSourceConfigurable2025}. We select the CV32E40PX core, extended with the eXtension Interface (XIF) to support custom co-processor integration.
The memory subsystem is organized into four regions: a 64~kB code bank, a 64~kB data bank, a 3.7~MB weight memory holding the network weights as per the quantization outlined in Section~\ref{subsec:quant}, and a $4 \times 8$~kB interleaved data bank, the latter enabling the conflict-free parallel access patterns.
The peripheral fabric is built on an \mbox{NtoM} bus interconnect and exposes a minimal set of base peripherals, including a DMA engine, power management, fast interrupt control, and interfaces for communication with the external world such as UART, JTAG, and SPI. Figure~\ref{fig:clk_comp}(a) shows, for the MCU baseline relying on the scalar core only, a pie-chart breakdown of clock cycles among the key computations done during the wake phase, highlighting that PW and DW convolutions dominate the overall footprint, which will thus be the focus of the next ablations in Fig.~\ref{fig:clk_comp}(b).

\subsubsection*{Matrix co-processor}

\begin{figure*}
    \centering
    \includegraphics[width=0.85\linewidth]{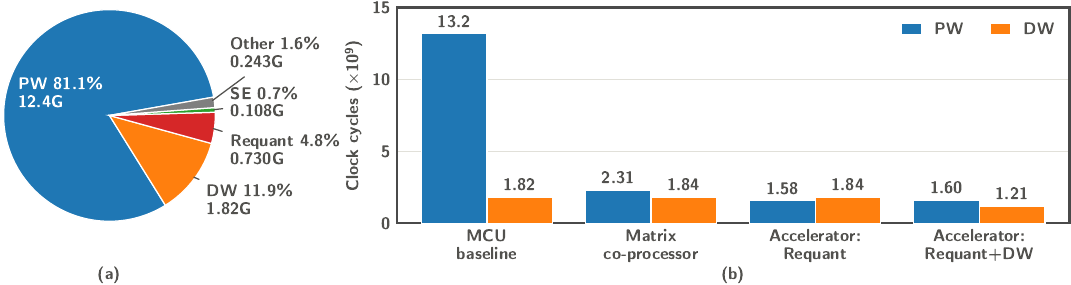}
    \vspace*{-1.5mm}
    \caption{Wake-phase clock cycles of the backbone per operation class. (a)~Clock cycles breakdown on the MCU baseline (scalar core only) into point-wise (PW) convolutions, depth-wise (DW) convolutions, squeeze-and-excitation (SE) layers, the per-channel requantization of the PW expansion outputs (Requant), and the remaining operations (Other). (b)~PW and DW cycles for the MCU baseline, the MCU with the Quadrilatero matrix co-processor, the accelerator executing the per-channel requantization, and the accelerator additionally executing the DW convolutions, respectively. Requantization is counted within PW. The SE layers and the remaining operations are not shown in (b): they shrink from 0.35\,G cycles on the MCU baseline to 0.11\,G with the co-processor and the accelerator, for a total of 2.92\,G cycles per sample in the rightmost configuration. The early backbone is executed depth-first~\cite{goetschalckx2019breaking} in order to minimize the activation footprint and fit within the memory budget available in Hypnos.}
    \vspace*{-1.5mm}
    \label{fig:clk_comp}
\end{figure*}

To tackle the bottleneck of dense, compute-dominant operations, we attach a matrix co-processor via the XIF. For this purpose, we select the Quadrilatero accelerator~\cite{cammarataQuadrilateroRISCVProgrammable2025} for its availability in open source and its streamlined matrix ISA extension in a RISC-V system. Quadrilatero is suitable to accelerate PW convolutions, as they apply a $1\times1$ kernel across the feature map: for each of the $H\cdot W$ spatial positions, it multiplies the $C_\text{out}\times C_\text{in}$ weight matrix by the $C_\text{in}$-dimensional channel vector. Stacking all positions, the layer collapses into a single dense matrix--matrix product of shape $(C_\text{out}\times C_\text{in})\times(C_\text{in}\times HW)$. Every weight is then reused across all $H\cdot W$ positions and every activation across all $C_\text{out}$ filters. This high data reuse keeps the array's processing elements saturated. Figure~\ref{fig:clk_comp}(b) shows that using the matrix co-processor leads to a reduction of cycles needed to execute PW convolutions by $5.7\times$ over the scalar-core-only MCU baseline.

\subsubsection*{Wake-Phase Accelerator Design}\label{ssubsec:acc}

\begin{figure}
    \centering
    \includegraphics[width=\linewidth]{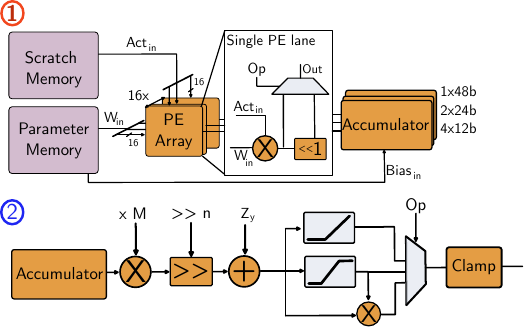}
    \caption{Accelerator datapath. {\color{red}\raisebox{.5pt}{\textcircled{\raisebox{-0.9pt}{1}}}} One of the 16 PE lanes: the activation word Act$_\text{in}$ and weight word W$_\text{in}$ are read as INT16, INT8, or INT4 operands according to Op, optionally doubled by a one-bit left shift ($\ll$1) for the distance mode, and accumulated into a 48-bit register preloaded through Bias$_\text{in}$. {\color{blue}\raisebox{.5pt}{\textcircled{\raisebox{-0.9pt}{2}}}} The shared post-MAC unit: multiply by $M$, shift by $n$ with a selectable rounding mode, add $z_y$, optional ReLU, hard-sigmoid, or hard-swish (through the second multiplier), and saturation to the output width.}
    \label{fig:blocks}
\end{figure}

The accelerator's main purpose is two-fold: (i)~accelerating vector-heavy operations (e.g.,~DW convolutions, per-channel requantization) with support for mixed INT4/INT8/INT16 operands, and (ii)~ensuring a fully on-chip wake phase, including PQ-related storage for samples encountered during daily operation and held compressed until the next sleep phase.
The accelerator architecture is shown in Fig.~\ref{fig:system} and its datapath is shown in Fig.~\ref{fig:blocks}, where DW convolution, requantization, and distance computation can be executed with a common template: a multiply--accumulate body followed by a per-channel post-operation. The PE array ({\color{red}\raisebox{.5pt}{\textcircled{\raisebox{-0.9pt}{1}}}}) has 16 lanes, each multiplying a 16-bit activation word read from the scratch memory (\texttt{Act$_\text{in}$}) with a 16-bit weight word (\texttt{W$_\text{in}$}) holding a DW kernel, an FC weight, a centroid, or a prototype, depending on the operation.
The mode signal \texttt{Op} sets how a word is read (i.e.~one INT16, two INT8, or four INT4 operands), so throughput follows the layer precision, from 16 to 64 multiply--accumulate (MAC) operations per cycle.
Products accumulate into a 48-bit register per lane (i.e.~one 48-bit, two 24-bit, or four 12-bit sums), preloaded through \texttt{Bias$_\text{in}$}.
The single post-MAC unit ({\color{blue}\raisebox{.5pt}{\textcircled{\raisebox{-0.9pt}{2}}}}), shared by all lanes, %
realizes the requantization of Eq.~\eqref{eq:requant}. 
Figure~\ref{fig:clk_comp}(b) shows that, compared to the MCU with matrix co-processor, offloading the per-channel requantization to the accelerator reduces the PW cycle count by $1.46\times$ (from 2.31\,G to 1.58\,G), and additionally offloading the DW convolutions reduces the DW cycle count by $1.52\times$ (from 1.84\,G to 1.21\,G), for a $1.46\times$ reduction of the full wake phase, from 4.26\,G to 2.92\,G cycles per sample.

\subsubsection*{Smart Charger}\label{ssubsec:charger}

Since it has access to wall-plug power, the implementation of the smart charger can be simplified to a software-based execution, taking place in three main steps.
First, the daily samples accumulated during the wake phase in the edge device (i.e.~PQ codes in their compressed form) are retrieved and added to the replay buffer maintained in the smart charger. The weights of the encoder, as well as the prototypes, are also retrieved for the next step.
Second, gradient-based consolidation retrains the encoder and the head, with samples retrieved from the replay buffer that were reconstructed via PQ decoding.
Third, consolidated parameters are written back to the edge device.
The main constraint for the smart charger is to have access to sufficient memory to store all accumulated samples across wake--sleep phases for the sleep replay buffer, as well as for executing full gradient-based learning on the encoder and head of the network, which here amounts to ${\sim}10$\,MB in total.

\section{Experimental Results}\label{sec:results}

In this section, we first describe the training and quantization pipelines of the deployed network (Section~\ref{ssec:training}) with the CORe50 \mbox{NICv2-391}~\cite{pellegriniLatentReplayRealTime2020} benchmark. Then, we validate the Hypnos device--charger system by prototyping it in a Xilinx ZCU104 FPGA (Section~\ref{ssec:implFPGA}). Finally, we report the post-synthesis area and power of the edge device implemented in a 28-nm CMOS technology node and compare Hypnos with the state of the art in Section~\ref{ssec:sota}.

\subsection{Network Training and Quantization}\label{ssec:training}

The deployed network is quantized offline to match the integer datapath of Section~\ref{ssubsec:acc}. 
Starting from a MobileNetV3-Large backbone pretrained on ImageNet, the
network is split into a frozen feature extractor and a plastic head at layer~15, the operating point defined in Section~\ref{subsec:quant}.
The quantization follows the mixed-precision scheme of Section~\ref{subsec:quant}, realized in Brevitas~\cite{pappalardoXilinxBrevitasRelease2025}, by combining post-training calibration and fine-tuning.
The scales are calibrated on ImageNet, the head is fine-tuned on the first
\mbox{NICv2-391} batch so that the classifier operates in the quantized
latent space, and a final calibration pass on CORe50 freezes all scales.
The model is then exported as an integer-only firmware bundle (INT8 weights, INT32 biases, per-channel requantization multipliers and zero points, and INT8 PQ centroids fitted through the quantized backbone). \par

\subsection{FPGA Implementation Results}\label{ssec:implFPGA}

For the FPGA validation of the full Hypnos device--charger system, we use a Xilinx ZCU104 board. The edge device (i.e.~the X-HEEP RISC-V-based MCU with the Quadrilatero co-processor and the wake-phase accelerator) is prototyped in the programmable logic (PL) in its INT8-activation variant, while a software-based implementation of the smart charger is done in the processing system (PS). The wake phase therefore takes place fully in the PL, while PL--PS (i.e.~device--charger) exchanges taking place only during the sleep phase. During the latter, the PYNQ subsystem abstraction is used so that the PS can access PL memories for retrieval and consolidation. The MCU firmware is programmed over JTAG through OpenOCD, the \mbox{NICv2-391} frames are streamed into the device over SPI, emulating a camera interface, and execution logs are collected over UART. 

The FPGA implementation fully validates the system with a final accuracy of 78.6\% and the number of cycles reported in Fig.~\ref{fig:clk_comp}. The PL resource utilization numbers are reported in Table~\ref{tab:resource_util}.
The per-block breakdown confirms the minimal wake-phase accelerator premise of Section~\ref{ssubsec:acc}: the accelerator occupies 18.8\,\% of the logic LUTs and 32 DSP blocks, 14.0\,\% of those used by the design, while the matrix co-processor takes 83.8\,\% of the DSPs.
The design is instead dominated by memory, with all of the device's URAM and 98.7\,\% of its BRAM consumed: the 3.7\,MB weight memory fills the 3.4\,MB of URAM and spills into BRAM, which further hosts the remaining MCU banks (756\,kB in total for the MCU baseline) and the scratch, codebook, prototype, and PQ-code memories of the accelerator (630\,kB). This inversion, inexpensive compute and dominant storage, reflects the design goal of the wake phase: keeping the full model state on chip is the cost that enables single-sample online adaptation, not the logic operating on it.
\begin{table}[t]
\centering
\caption{Post-implementation FPGA resource utilization of the Hypnos-A8 edge device on the ZCU104, broken down per block.}
\label{tab:resource_util}
\begin{threeparttable}
\footnotesize\setlength{\tabcolsep}{2.7pt}%
\newcommand*{\pct}[1]{{\scriptsize #1\%}}%
\begin{tabular}{@{}lrrrrrr@{}}
\toprule
Block\tnote{a} & LUTs & FFs & DSPs & LUTRAM\tnote{b} & BRAM\tnote{b} & URAM\tnote{b}      \\
\scriptsize(design relative \%) &  &   &      &        &      & \\ \midrule
MCU baseline & 22,573      & 16,132      & 5           & 169.0\,kB   & 756.0\,kB   & 3.4\,MB      \\
                   & \pct{26.0}  & \pct{29.1}  & \pct{2.2}   & \pct{40.2}  & \pct{54.5}  & \pct{100.0}  \\ \addlinespace
Matrix co-proc. & 23,049   & 10,320      & 192         & --          & --          & --           \\
                   & \pct{26.5}  & \pct{18.6}  & \pct{83.8}  &             &             &              \\ \addlinespace
Accelerator  & 16,334      & 8,877       & 32          & 225.0\,kB   & 612.0\,kB   & --           \\
                   & \pct{18.8}  & \pct{16.0}  & \pct{14.0}  & \pct{53.5}  & \pct{44.2}  &              \\ \addlinespace
Accelerator OBI & 8,076 & 404     & --          & 14.1\,kB    & 18.0\,kB    & --           \\
                   & \pct{9.3}  & \pct{0.7}   &             & \pct{3.3}   & \pct{1.3}   &              \\ \addlinespace
Others (system) & 16,897      & 19,781      & --          & 12.3\,kB    & --          & --           \\
                   & \pct{19.4} & \pct{35.6}  &             & \pct{2.9}   &             &              \\ \midrule
Total        & 86,920 & 55,514  & 229         & 420.4\,kB   & 1.4\,MB     & 3.4\,MB      \\
{\scriptsize(total ZCU104 device \%)} & \pct{37.7}  & \pct{12.0}  & \pct{13.3}  & \pct{52.9}  & \pct{98.7}  & \pct{100.0}  \\ \bottomrule%
\end{tabular}
\begin{tablenotes}\scriptsize
\item[a] Vivado 2023.1, \texttt{report\_utilization -hierarchical} on the routed checkpoint of implementation run \texttt{impl\_1} (XCZU7EV). Post-implementation figures for the A8 configuration (INT8 activations, 384\,kB scratch memory). Relative \%: share of the design total; total \%: share of the device.
\item[b] Capacity of the occupied primitives: 64\,bit per LUTRAM, 36/18\,kbit per RAMB36/RAMB18, 288\,kbit per URAM; 1\,kB${}={}$1024\,B.

\end{tablenotes}
\end{threeparttable}
\end{table}

\subsection{Comparison with the State of the Art}\label{ssec:sota}

To derive power and silicon area estimates, we synthesize the edge device in a 28-nm CMOS technology node. As different configurations lead to different accuracy--footprint tradeoffs (Section~\ref{sec:algo}), we synthesize Hypnos in both INT8- and INT16-activation variants, denoted respectively as A8 and A16. While this requires a larger scratchpad memory for activations (from 384\,kB to 743\,kB), the logic is otherwise unchanged as the accelerator's datapath already supports INT16 operands. As Table~\ref{tab:learning_cost} shows, the A16 variant thus leads to a higher wake-phase energy per sample (from 0.26\,J to 0.35\,J at 0.9\,V) and area (from 9.43\,mm$^2$ to 10.07\,mm$^2$), in exchange for a substantial accuracy boost (from 78.6\% to 81.6\%).

The approach of quantized-latent-replay-based continual learning (QLR-CL) proposed in~\cite{ravagliaTinyMLPlatformOnDevice2021} is, to the best of our knowledge, the only other work demonstrating CORe50 in a low-power edge device. As that work reports several configurations along the accuracy--footprint tradeoff, Table~\ref{tab:learning_cost} includes its two extreme points, both at the highest compression reported (INT8 weights, INT7 latent activations): the accuracy-oriented split at $l=19$ and the latency-oriented split at $l=27$. Both Hypnos and QLR-CL use the same \mbox{NICv2-391} protocol for CORe50.

\begin{table}[t]
\centering
\caption{Per-sample learning cost of Hypnos and QLR-CL on
\mbox{NICv2-391}. Hypnos is reported in two synthesized variants:
\textit{A8}, the FPGA-validated design with INT8 activations, and
\textit{A16}, with a larger scratch memory that fits INT16 activations.}
\label{tab:learning_cost}
\begin{threeparttable}
\scriptsize\setlength{\tabcolsep}{1.5pt}
\begin{tabular}{@{}l@{\hspace{4pt}}cccc@{}}
\toprule
 & Hypnos A8 & Hypnos A16 & QLR-CL $l{=}19$ & QLR-CL $l{=}27$ \\ \midrule
Backbone                       & MobileNetV3 & MobileNetV3 & MobileNetV1 & MobileNetV1 \\
Latent split (layer)           & 15 & 15 & 19 & 27 \\
Weight precision               & INT8\tnote{a} & INT8\tnote{a} & INT8\tnote{b} & INT8\tnote{b} \\
Activation precision           & INT8 & INT16 & INT8\tnote{b} & INT8\tnote{b} \\
Replay buffer precision        & PQ (8-bit) & PQ (8-bit) & INT7 & INT7 \\
Learning OPs/step\tnote{c}     & 0.343\,G & 0.343\,G & 7.70\,G & 5\,M \\
Learning steps (total)         & 741\,k & 741\,k & 2.80\,M & 2.80\,M \\
 \midrule
Weight memory                  & 3.70\,MB & 3.70\,MB & 9.10\,MB & 3.04\,MB \\
Total on-device memory\tnote{d} & 4.35\,MB & 4.71\,MB & 107.4\,MB & 6.04\,MB \\
Replay buffer\tnote{e}         & 5.48\,MB & 5.48\,MB & 98.3\,MB & 3\,MB \\ \midrule
Technology                     & 28\,nm & 28\,nm & 22\,nm & 22\,nm \\
Supply voltage                 & 0.9\,V & 0.9\,V & 0.6--0.8\,V & 0.6--0.8\,V \\
Clock frequency                & 375\,MHz & 375\,MHz & 375\,MHz & 375\,MHz \\
Area\tnote{f}                  & 9.43\,mm$^2$ & 10.07\,mm$^2$ & 12\,mm$^2$ & 12\,mm$^2$ \\
On-device latency              & 7.79\,s\tnote{g} & 10.54\,s\tnote{g} & 8.4\,s\tnote{h} & ${\sim}73$\,ms\tnote{h} \\
Sleep step latency\tnote{i}    & 231\,ms & 231\,ms & -- & -- \\
Power\tnote{f}                 & 33.3\,mW & 33.3\,mW & 62\,mW & 62\,mW \\
Wake energy\tnote{f}           & 0.26\,J & 0.35\,J & 0.51\,J & 0.43\,mJ \\
Fully on-chip\tnote{j}         & \textcolor{green!60!black}{\checkmark} & \textcolor{green!60!black}{\checkmark} & \textcolor{red}{$\times$} & \textcolor{red}{$\times$} \\ \midrule
\textbf{Accuracy (\%)}         & $\mathbf{78.6\pm1.2}$ & $\mathbf{81.6\pm2.1}$ & $\mathbf{75.2\pm1.1}$ & $\mathbf{67.2\pm1.1}$ \\ \bottomrule
\end{tabular}
\begin{tablenotes}\scriptsize
\item[a] INT4 for the fully-connected layers of the SE blocks.
\item[b] Frozen layers only; the plastic layers of QLR-CL operate in FP32.
\item[c] Operations of one learning step on one sample, executed on the charger for Hypnos and on the device for QLR-CL.
\item[d] Weights, activation scratch, PQ codes, prototypes, and codebooks for Hypnos. Weights plus replay buffer for QLR-CL.
\item[e] The QLR-CL buffer resides in external memory and must be accessed during operation. The Hypnos buffer is accessed only during the sleep phase, on the charger, where the on-chip constraint does not apply, while the PQ codes accumulated during the wake phase reside in the on-chip code buffer.
\item[f] Post-synthesis results of the complete edge device (MCU, matrix co-processor, wake-phase accelerator, and on-chip memories) in a 28-nm technology. QLR-CL figures as reported in~\cite{ravagliaTinyMLPlatformOnDevice2021}.
\item[g] Wake-phase execution of one sample. Cycle counts measured on the FPGA prototype and reported at the 375-MHz clock of the synthesized design.
\item[h] QLR-CL has no sleep phase: training runs on the device as each batch arrives and is included in this latency (0.352\,s per step at $l{=}19$; 62\,ms frozen $+$ 11\,ms training at $l{=}27$).
\item[i] Measured on the FPGA prototype's processing system CPU with PyTorch.
\item[j] No external memory accesses during the wake phase; verified for Hypnos through the FPGA prototype.
\end{tablenotes}
\end{threeparttable}
\end{table}

A learning step is one gradient update of the plastic head on a mini-batch. Over the same stream of training images, QLR-CL updates its head on every incoming batch and performs 2.80\,M steps in total, whereas Hypnos consolidates only during sleep and needs 741\,k, $3.8\times$ fewer. Each Hypnos step takes 231\,ms on the charger and therefore does not add to the latency experienced during daily operation; spread over the 117\,k training images that follow the initial batch, the sleep phase amounts to 1.47\,s per sample.
On the device, Hypnos A8 processes a sample in 7.79\,s of wake-phase execution, below the 8.4\,s of the accuracy-oriented $l{=}19$ configuration, whose latency consists entirely of on-device training. A16 takes 10.54\,s ($1.35\times$), as its INT16 activations halve the number of operands the accelerator's lanes process per cycle.
The memory footprints explain the fully on-chip row: Hypnos keeps 3.7\,MB of weights on chip and confines its 5.48\,MB replay buffer to the sleep phase, where the on-chip constraint does not apply. In contrast, QLR-CL at $l{=}19$ streams its 98.3\,MB buffer from external memory on every learning step.
In sum, QLR-CL at $l{=}27$ processes samples $107\times$ faster than Hypnos A8 but concedes 11.4\,pp of accuracy and still relies on external memory, while $l{=}19$ is slightly slower than A8, concedes 3.4\,pp, and keeps its buffer off-chip.
Hypnos A8 is the only system that sustains accuracy (78.6\,\%) while running the wake phase fully on-chip, and A16 raises it further to 81.6\,\% for the 0.64\,mm$^2$ of additional scratch memory and $1.35\times$ the wake-phase energy per sample of A8.
This deployment-level result mirrors the algorithmic analysis of Fig.~\ref{fig:flop_comp}: the dual-phase algorithm outperformed latent replay at every candidate split, and the co-designed platform carries that advantage through to the measured implementation.

\section{Conclusion}\label{sec:conclusion}

In this work, we propose Hypnos, a dual-phase CL platform that confines online adaptation to the edge device and memory consolidation to its charger, fully decoupling it from the cloud during its learning loop. 
This is achieved through three key insights: (i)~moving the sleep phase to a smart charger benefiting from extended memory and energy budgets, thereby binding the entire system to edge resource constraints, with no cloud-side assumption; (ii)~adjusting the memory--accuracy tradeoff by setting the latent split and the activation precision jointly, which trades 3.0\,pp of accuracy against 0.64\,mm$^2$ of scratch memory and $1.35\times$ the wake-phase energy per sample; and (iii)~efficiently splitting operations related to the frozen MobileNetV3 backbone, wake-phase adaptation, and PQ encoding into a low-footprint SoC consisting of a RISC-V MCU, a matrix co-processor, and a dedicated accelerator. 
The resulting open-source Hypnos platform was deployed on a Xilinx ZCU104 MPSoC, synthesized in a 28-nm technology at 9.43\,mm$^2$ and 33.3\,mW (10.07\,mm$^2$ for the INT16-activation variant), and benchmarked on the CORe50 \mbox{NICv2-391} protocol.
Hypnos A8 reaches 78.6\% accuracy at an on-device latency of 7.79\,s per sample at 375\,MHz, comparable to quantized latent replay at 3.4\,pp higher accuracy, with $3.8\times$ fewer learning steps and $22\times$ less learning compute per sample, while A16 reaches 81.6\% at 10.54\,s per sample. Among the systems compared, it is the only one that learns \textit{online} and \textit{fully on-chip} from real-image streams, within edge memory and power budgets and with no server in the loop. 

\section*{Acknowledgments}
The authors would like to thank Dr. Martin Lefebvre for his detailed input, feedback, and fruitful discussions, and Dr. Adrian Kneip for his help with the synthesis implementation.

\bibliography{tcas_bib}{}
\bibliographystyle{IEEEtran}

\vspace{11pt}

\begin{IEEEbiography}[{\includegraphics[width=1in,height=1.25in,clip,keepaspectratio]{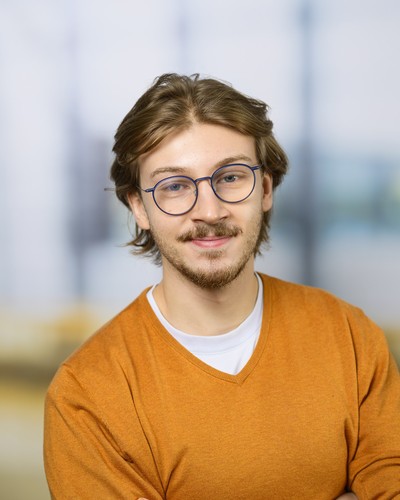}}]{Guilherme Guedes} (Graduate Student Member, IEEE) received the M.Sc. degree in electrical engineering from Faculdade de Engenharia da Universidade do Porto (FEUP), Porto, Portugal. He is currently pursuing the Ph.D. degree at Delft University of Technology (TU Delft), Delft, The Netherlands, under the supervision of Prof. C. Frenkel.

His current research interests include efficient continual learning algorithms and their implementation in silicon, and continual learning on temporal data.
\end{IEEEbiography}

\begin{IEEEbiography}[{\includegraphics[width=1in,height=1.25in,clip,keepaspectratio]{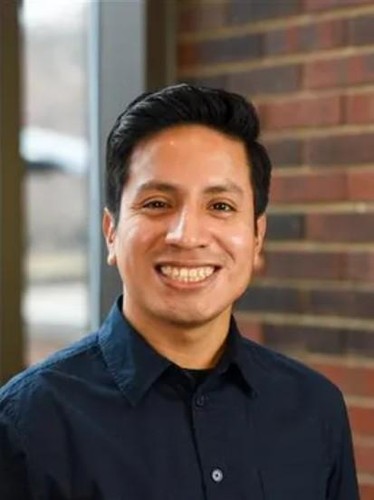}}]{Marco Paul E. Apolinario} (Member, IEEE) received the B.Sc. degree in electronics engineering from the National University of Engineering (UNI), Peru, in 2017, and the Ph.D. degree in electrical and computer engineering from Purdue University, USA, in 2025, under the supervision of Prof. Kaushik Roy.
He is currently a postdoctoral researcher in the Cognitive Sensor Nodes and Systems (CogSys) laboratory led by Prof. Frenkel at TU Delft, The Netherlands, where he was previously an NSF AccelNet NeuroPAC visiting researcher in 2024. His research centers on hardware-software co-design for neuro-inspired on-device intelligence, with a particular interest in memory-efficient and biologically inspired local learning rules, continual and in-context learning, and hardware-aware training for emerging substrates such as in-memory and neuromorphic computing.
Dr. Apolinario is a co-author of several research papers in IEEE and machine learning venues, receiving the Best Student Paper Award at the 2026 IEEE/CVF WACV. He served as a Review Committee Member for IEEE ISCAS 2025 and reviews for numerous IEEE journals and machine learning conferences, including TBioCAS, NeurIPS, and ICCV. He is also active in student mentoring and outreach.
\end{IEEEbiography}
\vfill
\newpage
\begin{IEEEbiography}[{\includegraphics[width=1.25in,height=1.3in,clip,keepaspectratio]{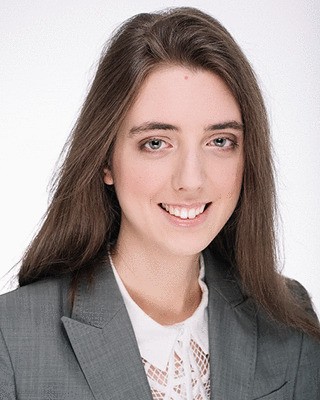}}]{Charlotte Frenkel}
(Senior Member, IEEE) received the M.Sc. degree (\textit{summa cum laude}) in Electromechanical Engineering and the Ph.D. degree in Engineering Science from Universit\'e catholique de Louvain (UCLouvain), Louvain-la-Neuve, Belgium in 2015 and 2020, respectively. In February 2020, she joined the Institute of Neuroinformatics, UZH and ETH Zurich, Switzerland, as a postdoctoral researcher. She is an Assistant Professor at Delft University of Technology, Delft, The Netherlands, since July 2022, and a Research Scientist at Google since February 2026.
 
Her research aims at bridging the bottom-up (bio-inspired) and top-down (engineering-driven) design approaches toward neuromorphic intelligence, with a focus on hardware-algorithm co-design for (Neuro)AI, digital hardware accelerators, and brain-inspired on-device learning.
 
Dr. Frenkel received a best paper award at the IEEE International Symposium on Circuits and Systems (ISCAS) 2020 conference in the \textit{Neural Networks} track, and her Ph.D. thesis was awarded the FNRS-FWO / Nokia Bell Scientific Award 2021 and the FNRS-FWO / IBM Innovation Award 2021. In 2023, she was awarded prestigious Veni and AiNed Fellowship grants from the Dutch Research Council (NWO). She presented several invited talks, including keynotes at the tinyML EMEA technical forum 2021 and at the Neuro-Inspired Computational Elements (NICE) neuromorphic conference 2021. She serves or has served as a program co-chair of NICE 2023-2024 and of the tinyML Research Symposium 2024, as a TPC member of IEEE ISSCC for 2027 and IEEE ESSERC for 2022-2024, and as an associate editor for the IEEE Transactions on Biomedical Circuits and Systems for 2022-2025.
\end{IEEEbiography}

\vfill

\end{document}